\documentclass[sigconf,nonacm]{acmart}
\AtBeginDocument{%
  }

\setcopyright{none}
\newcommand{\keff}{\ensuremath{K_{\mathrm{eff}}}}
\newcommand{\frad}{\ensuremath{F_{\mathrm{rad}}}}
\newcommand{\fax}{\ensuremath{F_{\mathrm{ax}}}}

\usepackage{bm}

\usepackage{graphicx}
\usepackage{subcaption}
\usepackage{booktabs}
\graphicspath{{figures/}}

\begin{document}

\title{Analysis of Federated Graph Neural Networks for Predicting Nuclear Core Properties}

\settopmatter{authorsperrow=4}
\author{Oliver Kožul}
\email{oliver.kozul@dmi.uns.ac.rs}
\affiliation{%
	\institution{Faculty of Sciences, University of Novi Sad}
	\city{Novi Sad}
	\country{Serbia}}

\author{Mihailo Ilić}
\email{milic@dmi.uns.ac.rs}
\affiliation{%
	\institution{Faculty of Sciences, University of Novi Sad}
	\city{Novi Sad}
	\country{Serbia}}

\author{Julia Bartos}
\email{julia.bartos@nrgpallas.com}
\affiliation{%
	\institution{NRG PALLAS}
	\city{Petten}
	\country{Netherlands}}

\author{Tuan Tran}
\email{tuan.tran@nrgpallas.com}
\affiliation{%
	\institution{NRG PALLAS}
	\city{Petten}
	\country{Netherlands}}

\author{Miloš Savić}
\email{svc@dmi.uns.ac.rs}
\affiliation{%
	\institution{Faculty of Sciences, University of Novi Sad}
	\city{Novi Sad}
	\country{Serbia}}

\author{Vladimir Kurbalija}
\email{kurba@dmi.uns.ac.rs}
\affiliation{%
	\institution{Faculty of Sciences, University of Novi Sad}
	\city{Novi Sad}
	\country{Serbia}}

\author{Edo Frederix}
\email{edo.frederix@nrgpallas.com}
\affiliation{%
	\institution{NRG PALLAS}
	\city{Petten}
	\country{Netherlands}}

\author{Dusan Jakovetic}
\email{dusan.jakovetic@dmi.uns.ac.rs}
\affiliation{%
	\institution{Faculty of Sciences, University of Novi Sad}
	\city{Novi Sad}
	\country{Serbia}}

\renewcommand{\shortauthors}{Kožul et al.}

\begin{abstract}
Federated learning has grown in popularity owing to the increasing data privacy and safety concerns. It allows multiple parties to collaboratively train a model without exchanging their raw data. As an example of highly sensitive infrastructure of national importance, nuclear reactors benefit from this approach, where---while pooling multiple large datasets from different institutions could improve machine learning solutions---the confidentiality of the datasets is paramount. In this paper, we demonstrate that a graph neural network (GNN) outperforms the convolutional neural network (CNN) baseline in a federated setting for a use case concerning nuclear reactor core optimization. The GNN architecture is better suited for the nuclear reactor geometry as, unlike the CNN, it does not require padding the cylindrical core and does not assume translation invariance. The key finding of this work is that input representation is more important than model size and complexity. More sophisticated federated algorithms were employed, but the improvements derived were of lower magnitude. Experimenting with training corpus volume led to the discovery that greater fuel inventory variance is more beneficial than arrangements of existing cases, which is a direct consequence of data heterogeneity. Going forward, focus will remain on enhancing the input and readout representation. 
\end{abstract}



\maketitle

\section{Introduction}
\label{sec:intro}

Federated learning (FL) is a distributed machine learning paradigm where multiple clients collaboratively train a model while keeping their data local~\cite{mcmahan2017communication,kairouz2021advances}. During each training round, the central server relays the current global model to the clients, each client trains its model on its local data, after which the server aggregates the clients' returned updates into a new global model, most commonly via the FedAvg algorithm~\cite{mcmahan2017communication}. 

Although the motivation for distributed learning techniques is usually efficiency, FL is predominantly driven by privacy, as in medicine, where hospitals can jointly train diagnostic models without exchanging private information such as patient records~\cite{rieke2020future}. A similar sentiment is shared in the nuclear field, where power plants  work with similar, but confidential, datasets, including their own fuel inventory. By employing FL algorithms, multiple reactors could create a global model using pooled knowledge while keeping their data decentralized, an approach referred to as cross-silo FL~\cite{kairouz2021advances}. Through this collaboration, the global model is exposed to a greater variety of fuel compositions and arrangements, as well as reactor geometries and properties, creating a robust solution for this task. 

The main challenge of this setting is the heterogeneity of data. Since each site has different fuel inventories, the client data is not necessarily independent and identically distributed (non-IID). Due to the lack of data uniformity, the optimum of each client's objective might not align with the global objective, and local training gradually pulls the client weights apart. This phenomenon is known as client drift, and it can negatively affect global model performance due to client outputs canceling each other out through disagreement~\cite{zhao2018noniid,karimireddy2020scaffold,ye2023heterogeneous}. 

The use case of our study focuses on the first step in reactor core optimization, predicting key core parameters, based on fuel loading. The reactor core contains fuel assemblies, which for this use case are represented by the type and total weight of the fissile material in the fuel assembly. The output of interest is the so-called effective multiplication factor (\keff{}), which is a global parameter that describes the neutron balance in the reactor core. 

Although multiple model architectures were tested for this use case, graph neural networks (GNNs) yielded the best results. This is primarily due to the core structure and geometry of the input. A loading pattern assigns fuel assemblies to positions on a circularly truncated square lattice in which every position has a neighborhood. A natural representation of such relationships is through a graph, with neighboring nodes being connected by edges. A GNN exchanges information only along the edges of its graph which approximately mirrors the local coupling between assemblies~\cite{bronstein2021geometric}.

The results reported in this paper focus on federated GNN surrogates that were used to predict the effective multiplication factor \keff{} computed by a neutronics solver. Whereas the work relies on well-established federated and graph methods, its novelty stems from an empirical analysis of FL limitations on heterogeneous reactor data:
\begin{enumerate}
	\item To the best of our knowledge, the first study of federated GNN surrogates of a reactor physics solver, using an efficient model with 26{,}564 parameters trained across ten federated clients. 
	\item Evidence that input representation carries drastically greater importance compared to the size and sophistication of the federated model. 
	\item An ablation of federated optimization algorithms and enhancement of input representation, in which proximal regularization (FedProx), server-side adaptive optimization (Fed\-Adam), and physics channels jointly improve the \keff{} $R^2$ value by $+0.4258$. 
	\item Clear relationships between model performance and training data. Additionally, it was established that doubling the number of fuel compositions raises $R^2$ about eight times more than doubling the number of arrangements per composition, which bolsters efficiency of future data generation.
\end{enumerate}

\section{Related Work}
\label{sec:related}

\subsection{Graph Neural Networks}

Most modern GNNs follow the same principle where nodes exchange messages with their neighbors~\cite{gilmer2017neural,battaglia2018relational}. In every layer of the network, each node computes a message from its features and relays it to its neighbors that combine it with their own state. This procedure allows a network with $L$ layers to propagate information across $L$ hops of the graph. Graph convolutional networks (GCNs)~\cite{kipf2017gcn} approach this design with a normalized sum over the neighbors, with every neighbor carrying the same weight in the sum. GraphConv~\cite{morris2019wl} iterates upon this idea by applying different weight matrices to a node and to its neighbors, while polynomial spectral filters~\cite{defferrard2016chebnet} allow a single layer to reach a $k$-hop neighborhood. 

Owing to their attributes, GNNs are becoming a common choice for surrogates of physical simulators, with various applications such as particle dynamics~\cite{sanchez2020learning} and mesh-based solvers~\cite{pfaff2021mesh}.

\subsection{Federated Optimization under Heterogeneity}

FedAvg~\cite{mcmahan2017communication} is the default federated algorithm choice and functions well as a general-purpose approach. Its accuracy decreases as the client distributions diverge~\cite{zhao2018noniid}, which occurs with heterogeneous datasets. This client objective divergence can be combated client-side or server-side. FedProx~\cite{li2020fedprox} adds a proximal term to the local client objective that penalizes the distance between the local and the global weights, whereas SCAFFOLD~\cite{karimireddy2020scaffold} corrects the local updates via variance reduction. Adaptive federated optimization~\cite{reddi2021adaptive} is a server-side algorithm that treats the averaged client update as a pseudo-gradient and applies an adaptive optimizer such as Adam to the global model, which introduces momentum across rounds to accelerate the training process. 

Surveys of heterogeneity in federated learning~\cite{kairouz2021advances,li2020flchallenges,ye2023heterogeneous} distinguish several types of non-IID data, two of which are present in this work. Feature distribution skew is a product of clients handling different regions of the input space, while concept shift occurs when the same input maps to different outputs on different clients. Both are present in this task since clients work on different sets of fuel inventories, and fuel mass is not the only factor influencing core reactivity. 

\subsection{Federated Learning on Graphs}

Federated graph learning can be categorized by what each client represents: entire graphs, subgraphs, or nodes of a shared graph~\cite{he2021fedgraphnn,fu2022fgml}. The task described in this paper belongs to the first category. Xie et al.~\cite{xie2021gcfl} observed that graphs held by different clients tend to differ in both structure and features, and proposed clustering the clients based on the similarity of their gradient sequences. All graphs share the same structure in the model presented in this work, with the heterogeneity originating from node features. 

Although existing literature does touch on the topic of using neural-network surrogates for predicting core parameters such as \keff{}~\cite{zhang2026ai}, to the best of our knowledge, no work has been conducted on federated GNN surrogates of a reactor physics solver. 

\section{Data}
\label{sec:data}

The data used in this project has a two-level structure. A case represents a given inventory of fuel assembly types, represented by their fissile material content. A row is one
arrangement of that inventory within the core, otherwise referred to as core loading. Each case consists of 1{,}000 distinct arrangements.

\begin{figure}[t]
	\centering
	\includegraphics[width=\columnwidth]{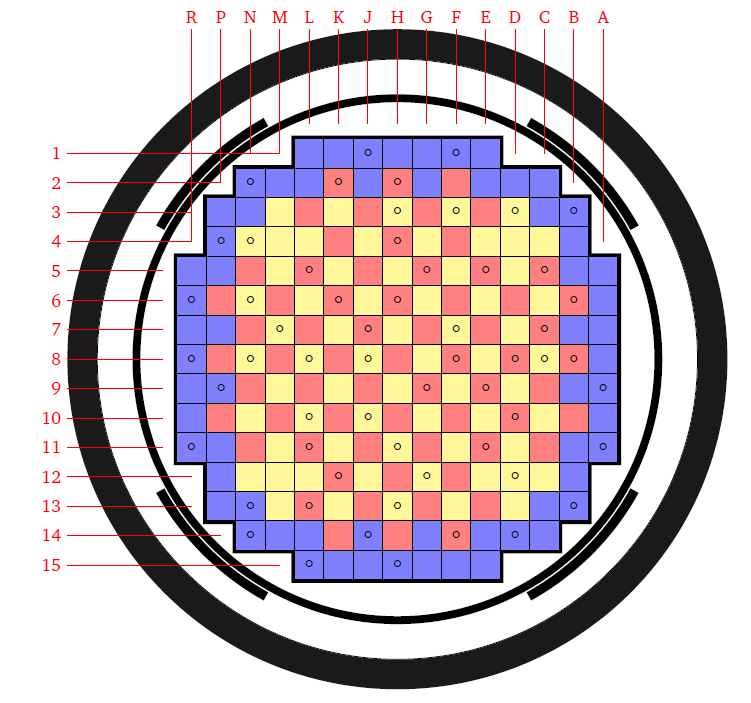}
	\caption{The BEAVRS benchmark core loading, with red and yellow showing the fuel assemblies and blue the reflector blocks.}
    \Description{An image depicting the BEAVRS benchmark core loading, with red and yellow showing the fuel assemblies and blue the reflector blocks.}
	\label{fig:BEAVRS}
\end{figure}

In this study, the BEAVRS pressurized water reactor benchmark was selected as a study case because it provides a high-fidelity reactor model and comprehensive documentation while remaining sufficiently simple for large-scale dataset generation. The core geometry is derived from the BEAVRS benchmark~\cite{horelik2013beavrs} and is shown in Figure~\ref{fig:BEAVRS}. It contains 193 fuel
positions surrounded by 48 reflector positions, amounting to 241 positions of interest in total. The lattice is four-connected, producing 448 edges. 

The model is designed to be exactly invariant under the eight symmetry operations of the
lattice. The fuel arrangements within a case are random shuffles and are therefore not symmetric, whereas the symmetry of the core hardware could not be assumed and had to be tested. Since the
arrangements within a case are random shuffles, the mean flux field of each case must inherit
the symmetry of the fixed core hardware. Compared against an empirical split-half baseline,
the asymmetry of the mean flux field has a ratio of 1.2 to 1.3 across all seven non-identity
operations, which corresponds to a systematic component equal to $\approx$4\% of the spatial
variation of the flux field, indicating that the core hardware is symmetric to a good approximation. As a control, the same test applied to the mean loading gives a
ratio of 1.41, which matches the $\sqrt{2}$ expected for uncorrelated fields and confirms that
the arrangements are exchangeable.

A neutronics solver (i.e., the MGRAC multi-group nodal diffusion solver within the OSCAR-5 code system \cite{muller1994development}) maps a loading arrangement to a set of computed core parameters, and fast surrogates of this mapping are an established tool in reactor physics~\cite{zhang2026ai}. The goal of the learning task is to best approximate the desired output parameters for a given arrangement. The inputs are 193 fuel-assembly masses of an arrangement, while the outputs are three quantities, respectively denoted by \keff{}, \frad{}, and \fax{}. Throughout this work, we focus exclusively on \keff{}, the figure of primary interest for this use case, while \frad{} and \fax{} are to be examined in future work. 

\begin{figure*}[t]
	\centering
	\includegraphics[width=\textwidth]{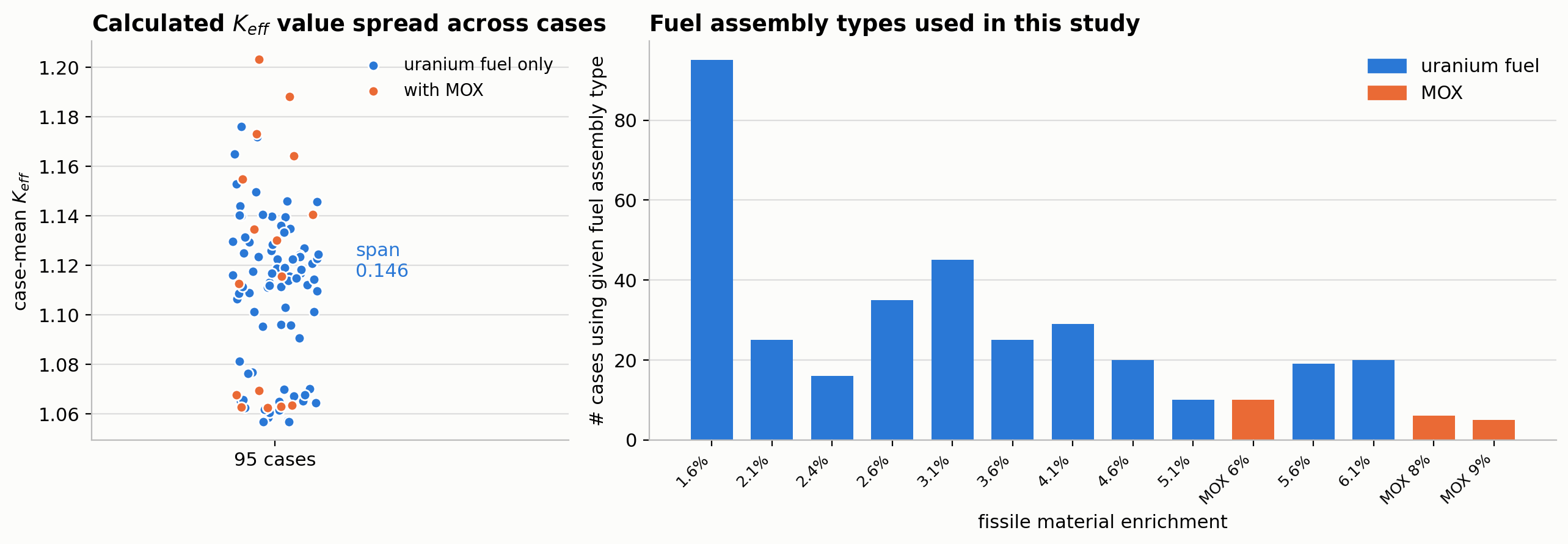}
	\caption{Calculated \keff{} value spread and fuel distribution across cases.}
	\Description{A figure that contains a scatterplot and bar chart showing the \keff{} spread and fuel distribution.}
	\label{fig:datachange}
\end{figure*}

The data contained 95 cases spanning 14 fuel levels (defined by the fissile material enrichment), three of which are mixed-oxide (MOX), with the rest containing only uranium as fissile material. Each case is part of a family (a subset of cases with similar properties), with each family containing around 10 consecutive cases (1 to 10, 11 to 20, $\ldots$). After removing rows with unrealistic entries, 94{,}988 rows were available for model training. \autoref{fig:datachange} presents a visualization of \keff{} variation by case followed by a fuel distribution breakdown, while a variance decomposition of \keff{} over the 95 cases is displayed in \autoref{tab:variance}.

\begin{table}[t]
	\centering
	\small
	\caption{Variance decomposition of \keff{}.}
	\Description{A table that provides a variance decomposition of \keff{}.}
	\label{tab:variance}
	\begin{tabular}{lrr}
		\toprule
		\textbf{Component} & \textbf{Variance} & \textbf{Share} \\
		\midrule
		Across cases (which inventory is loaded) & $1.142\times10^{-3}$ & 87.8\% \\
		Within cases (how it is arranged)        & $1.583\times10^{-4}$ & 12.2\% \\
		\midrule
		Total                                    & $1.300\times10^{-3}$ & 100\% \\
		\bottomrule
	\end{tabular}
\end{table}

\section{Methodology}
\label{sec:method}

\subsection{The Graph Neural Network}
\label{sec:gnn}

The model developed for this task is a graph convolutional network~\cite{kipf2017gcn} over the 241-node lattice:
three layers of width 64, each applying separate weight matrices to a node and to its
neighbors, extended with a second-order polynomial term, followed by layer normalization,
a rectifier, and dropout at 0.05. 

Each loading pattern of the analyzed dataset is converted to a graph with one node per lattice position: 193 fuel assemblies and 48 reflector blocks. Two nodes of the graph are connected if their positions within the lattice share a side, leading to a total of 448 edges. 

The readout concatenates the mean, standard deviation, and
maximum over all nodes and maps the result to the three target outputs. A node-level head predicts the
flux field during training only---an architectural choice that improved evaluation metrics as well. The final model that is described in this paper consists of 26{,}564 parameters, making it effective and compact.

The baseline is a convolutional network (CNN) using $3\times3$ kernels and group normalization, which the GNN models were designed to outperform. The CNN was more complex than our headline model, containing 56{,}547 parameters. One of the issues arising from this architectural choice becomes apparent when observing the core's geometry. The core is cylindrical and must be padded into a $17\times17$ rectangle for a convolution to apply, which wastes roughly a sixth of the grid on positions that do not exist. Nonetheless, the CNN was a logical choice for a baseline model due to the geometric interdependencies among neighboring inputs.  

Although the need for padding is a considerable CNN shortcoming, the main argument against this architecture is the assumption of translation invariance. In a reactor core, the distance from a boundary is a geometric feature that must not be undermined. A model like the CNN completely disregards this fact and neglects the physical interaction of inputs due to its translation-invariance assumption. This reasoning motivated the move to a graph network where there is finer control of input connectivity.

\subsection{Evaluation Protocol}
\label{sec:protocols}

The evaluation protocol used to report results of the frontier model divided the dataset into three groups. The method relied on a fixed seed when determining the case indices for the train--evaluation split.
Each of the three folds trains on 63 or 64 cases and is evaluated on the remaining 31 or 32, with every case being held out exactly once. In the federated configuration, the training cases are distributed round-robin across ten clients, contributing six or seven cases to each client. 

Since this evaluation process shuffles the case indices at random, it is possible for the training set not to include any cases of a given family. This is a challenging task for the model as it encounters examples of an unknown type during evaluation. Regardless, the model generalizes well and does not fall short compared to a model that had access to a portion of all case families during training. 

\section{Results}
\label{sec:results}

\begin{figure}[t]
	\centering
	\includegraphics[width=\columnwidth]{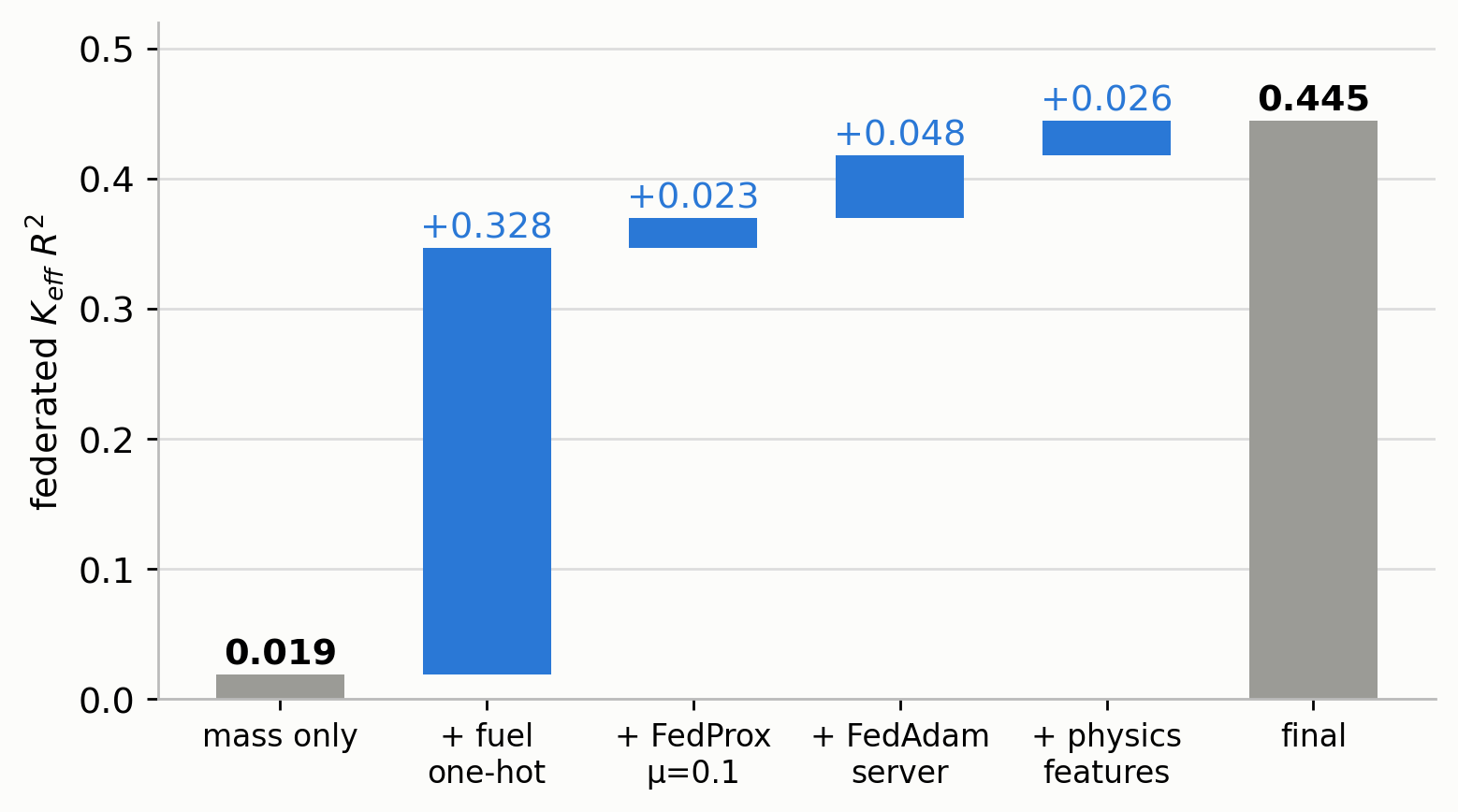}
	\caption{The ablation ladder, federated, 95 cases, grouped three-fold, per-case $R^2$. The one-hot channels were later superseded by the physics channels.}
	\Description{An ablation ladder that breaks down the model improvements by change made in the input representation and addition of federated algorithms.}
	\label{fig:ladder}
\end{figure}

All the results reported in this section are derived using the grouped three-fold protocol presented in Section~\ref{sec:protocols}, scored per case. A visualization of the incremental improvement of the GNN through the change of input representation and federated algorithms described in this section is presented in the form of an ablation ladder in \autoref{fig:ladder}. 

A comparison of model $R^2$ is shown in \autoref{tab:fedgap}. The addition of more sophisticated federated algorithms and improvement of data representation managed to bridge the gap between the federated and centralized approaches, but did not raise the centralized ceiling. Unlike the categorical channel approach, the continuous representation of fuels can be averaged across clients, rendering it the more federation-friendly option. 

In physical units, the final federated model reports a mean absolute error (MAE) of 708\,pcm (1\,pcm $=10^{-5}$). In its current state, the surrogate cannot replace the solver as MAE $\le$ 100\,pcm is acceptable, while MAE $\le$ 10\,pcm is ideal. 

\begin{table}[b]
	\centering
	\small
	\caption{Federated--centralized gap by input representation.}
	\Description{A table that shows the gap between the centralized and federated models.}
	\label{tab:fedgap}
	\begin{tabular}{lrrr}
		\toprule
		\textbf{Inputs} & \textbf{Centralized} & \textbf{Federated} & \textbf{Gap} \\
		\midrule
		Mass only              & 0.4202 & 0.0187 & $0.40$ \\
		+ Fuel-type one-hot    & 0.4165 & 0.4182 & $\approx 0.00$ (raw) \\
		+ Physics channels     & 0.4481 & 0.4445 & $\mathbf{0.004}$ \\
		\bottomrule
	\end{tabular}
	
	\footnotesize
	\parbox{\columnwidth}{\textit{Note.} FedProx and FedAdam were used when generating the one-hot and physics channel results.}
\end{table}

Centrally, the same features achieve only a $+0.028$ improvement over the mass-only inputs, with one-hot actually underperforming the naive approach. 

\subsection{Mass-Only Inputs in a Federated Setting}

The dataset consists of fuel masses for each node in the arrangement. The baseline result of our proposed model utilizes only standardized mass values.

The centralized variant reaches a \keff{} $R^2$ value of 0.4202 with mass-only inputs. Cases 60 to 66 score between $-0.06$ and $-0.33$ even though they contain three distinct fuel levels, none of which are mixed-oxide. The introduction of new information related to fuel types makes the prediction task of the mass-only model difficult, but does not explain the failure of these cases.

The federated counterpart of the above scores only $0.0187$, with two of three folds negative ($0.2634$, $-0.1131$, $-0.0979$), and 13 of 95 cases fall below 0.15. A federated--centralized gap of 0.40 is substantial. The disparity between approaches was first believed to derive from under-convergence, but that theory was ruled out quickly since round counts higher than 150 did not benefit the model. 

The federated approach was lackluster in this scenario due to the heterogeneity of the new dataset~\cite{ye2023heterogeneous} combined with an inadequate input representation. With mass-only inputs, each client learns a mapping from uranium mass to reactivity that is valid for the fuel families it holds. For a different fuel type, the same mass values can produce a vastly different outcome, leading to inconsistent results across cases. Averaging the contradictory mappings of clients destroys the signal, which the centralized model avoided because it had access to all the data at training time. 

\subsection{Fuel-Type Channels}

Two encodings of fuel identity were attempted in addition to the raw mass levels, and one yielded better results. The 95 cases draw from 14 discrete mass levels, of which each case utilizes three to five in total.

The first approach used a 14-way indicator per assembly. In the federated setting, this raises \keff{} $R^2$ from 0.019 to 0.347, the largest single improvement in the project at $+0.328$. The new feature brings cases 60 to 66 from negative to $+0.27$--$0.35$. This encoding was later superseded by the physics channels. Although effective, this addition does not resolve all of the problematic loadings. For example, case 91 incorporates an underrepresented fuel level, leading to a lack of training, and thus worse results. When held out, the case collapses to $-3.22$ centrally. This discovery motivated the search for a more robust solution that can capture fuel levels across all cases.  

\begin{figure}[t]
	\centering
	\includegraphics[width=\columnwidth]{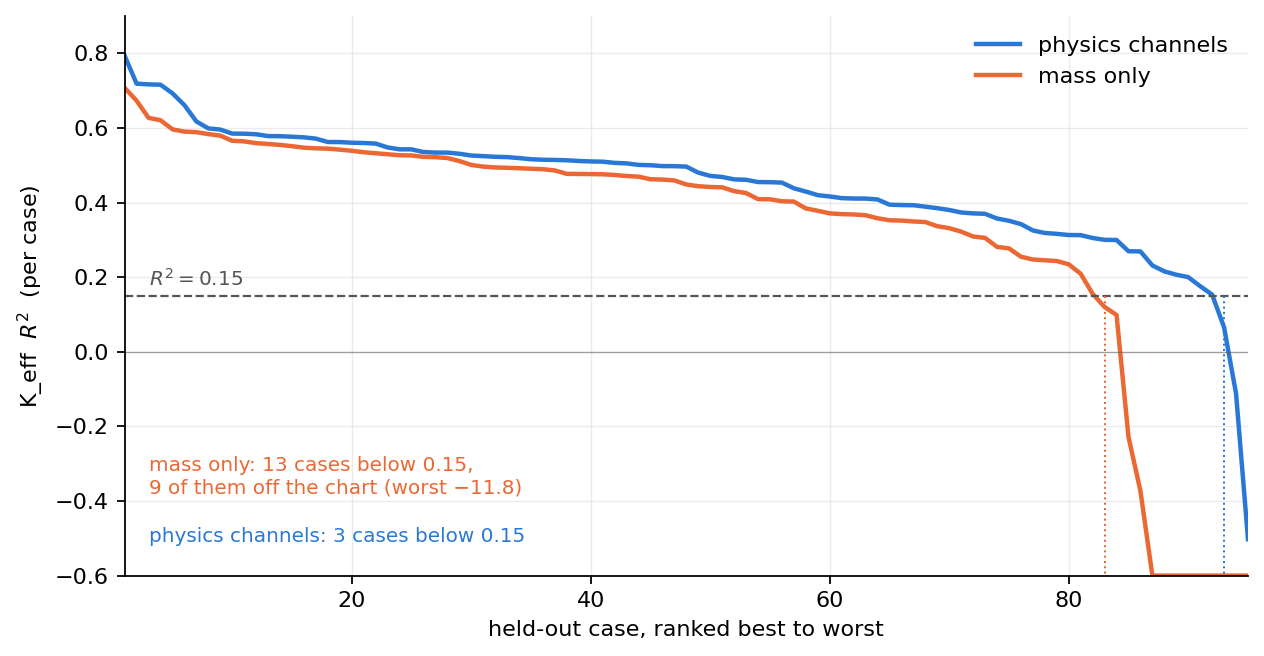}
	\caption{All 95 held-out cases ranked, mass-only against physics channels. The number of cases below the
		0.15 threshold declines from 13 to 3.}
	\Description{A graph visualizing the \keff{} $R^2$ value for each case, focusing on the number of cases below the 0.15 threshold.}
	\label{fig:percase}
\end{figure}

The superior method uses two columns per assembly: an enrichment proxy, linear in the mass level, and a binary mixed-oxide indicator. Centrally, this strategy scores 0.4481 against one-hot's 0.4165, and case 91 shifts from $-3.22$ to $+0.156$. The explanation of the improvement in results is simple. In the previous approach, the lack of data for some fuel levels led to the model failing. The new mechanism allows for interpolation between levels, which especially aids in prediction of underrepresented fuel levels. While not directly improving outputs for easier cases, this change increases system robustness, which is crucial when tackling outlier cases. In the federated setting, the number of failing cases decreases by 10, as shown in \autoref{fig:percase}. 

Two federated algorithms were added to the training loop in an attempt to improve results. The gains were modest, $+0.071$ in total compared to $+0.354$ introduced by the newly added input features. The algorithms in question are discussed in the following sections. 

\subsection{Proximal Regularization}

FedProx~\cite{li2020fedprox} is a client-side algorithm that adds a penalty $\tfrac{\mu}{2}\lVert \bm{w} - \bm{w}_{\text{global}} \rVert^2$ to each client's objective, discouraging the client from straying from the global weights within a round. A sweep was conducted to find the optimal value for the coefficient, with $\mu = 0.1$ yielding the greatest improvements. 

The goal of this term was to restrain the model from leaning heavily into rare features and arriving at a local optimum, which would degrade global performance, otherwise known as client drift. At $\mu = 0.1$ the gain is concentrated on the most heterogeneous fold. It raises the score by $+0.074$ on fold three, while the easiest fold experiences a loss of $0.009$. The penalty introduced by FedProx also suppresses legitimate client disagreement which explains the small loss on the easiest fold. 

\subsection{Server-Side Adaptive Optimization}

FedAdam~\cite{reddi2021adaptive} is a server-side federated algorithm that treats the aggregated client update as a pseudo-gradient and applies Adam to the global model. A learning rate of 0.01 and $\tau = 10^{-3}$ were used to raise the \keff{} $R^2$ value from $0.370$ to $0.4182$, an improvement of $+0.048$, with every fold up. Additionally, this algorithm change reduced the number of cases scoring below 0.15 from 12 to 5. The two easiest folds gain the most, consistent with a faster optimizer. Faster convergence also lets under-trained levels reach useful values, allowing more cases to clear the 0.15 threshold. The learning rate was not swept, the value of 0.01 was taken as a default and was not altered in future runs. 

\subsection{Learning Curves}

Although implementing different algorithms yielded improvements compared to the baseline model, feature representation proved the most important factor. Other than features and model complexity, it was critical to determine the value of an extended training corpus when evaluating the model. Two learning curves were measured, subsampling only after the fold split so that the held-out set is identical at every point. 

\begin{table}[t]
	\centering
	\footnotesize
	\caption{Learning curves, per-case $R^2$ on the same held-out cases at every point. }
	\label{tab:curves}
	\begin{subtable}[t]{0.47\columnwidth}
		\centering
		\begin{tabular}{rr}
			\toprule
			\textbf{Rows per case} & \textbf{\keff{} $R^2$} \\
			\midrule
			50   & 0.2668 \\
			100  & 0.3888 \\
			250  & 0.3828 \\
			500  & 0.4308 \\
			1000 & \textbf{0.4445} \\
			\bottomrule
		\end{tabular}
		\caption{Arrangements per case, all cases kept.}
		\Description{A table presenting the increase in \keff{} $R^2$ by increasing the number of arrangements used per case during model training.}
		\label{tab:curves-rows}
	\end{subtable}
	\hfill
	\begin{subtable}[t]{0.47\columnwidth}
		\centering
		\begin{tabular}{rr}
			\toprule
			\textbf{Training cases} & \textbf{\keff{} $R^2$} \\
			\midrule
			8  & $-2.771$ \\
			16 & $+0.026$ \\
			32 & $+0.287$\\
			48 & $+0.390$\\
			63 & $\bm{+0.445}$ \\
			\bottomrule
		\end{tabular}
		\caption{Training cases, all arrangements kept.}
		\Description{A table presenting the increase in \keff{} $R^2$ by increasing the number of cases used during model training.}
		\label{tab:curves-cases}
	\end{subtable}
\end{table}

\autoref{tab:curves} presents two learning curves related to sample size. Doubling the number of arrangements within a case increases the $R^2$ by $0.017$, whereas approximately $0.14$ is attained by doubling the total number of training cases. These values are slopes rather than differences between rows. By extrapolating from the reported values, reaching $R^2 = 0.60$ by adding arrangements alone would require around 600 times the current budget. Doubling the case count from 95 to roughly 190 is projected to have a similar effect. 

Neither projection has been tested, and extrapolation beyond the measured range is unreliable. The slope is decaying, making the $+0.14$ figure optimistic, while $+0.10$ seems to be the more conservative and realistic alternative. If more training data is requested in the future, it should be in the form of new fuel compositions rather than more arrangements of existing cases.

\section{Conclusion and Future Work}
\label{sec:conclusion}

FL is a powerful tool with a wide range of applications; combined with GNNs, it yields a robust solution that maintains the privacy of all institutions involved. In this paper, we tackle the issue of heterogeneous data employing techniques described in previous research~\cite{li2020fedprox,reddi2021adaptive,karimireddy2020scaffold,ye2023heterogeneous,kairouz2021advances,li2020flchallenges}. Adapting federated learning to this specific task proved challenging, yet the distributed nature of the approach is essential when handling data of this type. 

One of the central findings is that the most impactful changes for the federated model yielded neutral or slightly positive results for the centralized variant. This is to be expected since certain fuel levels were underrepresented in the dataset, which caused client drift. All the improvements described in previous sections were required to equalize the two techniques since the additional expressivity did not benefit the centralized approach, but it managed to raise the ceiling of the federated method required by the use case. 

In future work, we aim to further improve upon the input and readout representation. All results of our research point towards expressivity rather than model size and complexity. A richer readout containing additional parameters alongside the existing mean, maximum, and standard deviation is a top priority. Observing other core geometry patterns that do not violate the confirmed symmetry is also a promising lead. 

\begin{acks}
This work has been supported by the TURING project, funded by the European Union under the Horizon Europe programme, Grant Agreement No. 101215032. 
\end{acks}

\bibliographystyle{ACM-Reference-Format}
\bibliography{references}

\end{document}